\documentclass[2pt]{article}
\usepackage{amsmath}
\usepackage{amssymb}

\date{}

\title{Towards violation of Born's rule: description of a simple experiment}

\author{Andrei Khrennikov\\International Center for Mathematical Modelling
\\in Physics and Cognitive Sciences,\\
Linnaeus University, V\"axj\"o, S-35195, Sweden\\
Andrei.Khrennikov@lnu.se}

\begin{document}

\maketitle

\abstract{Recently   a new model with hidden variables of the wave type was elaborated, so called 
prequantum classical statistical field theory (PCSFT). Roughly speaking PCSFT is a classical signal theory 
applied to a special class of signals -- ``quantum systems''. PCSFT reproduces successfully all 
probabilistic predictions of QM, including correlations for entangled systems. This model peacefully coexists
with all known  no-go theorems, including Bell's theorem. In our approach QM is an approximate model.
All probabilistic predictions of QM are only (quite good) approximations of 
``real physical averages''. The latter are averages with respect to fluctuations of prequantum fields. In particular,
Born's rule is only an approximate rule. More precise experiments should demonstrate its violation. We present a simple 
experiment which has to produce statistical data 
violating Born's rule.  Since the PCSFT-presentation of this experiment may be difficult for experimenters, 
we reformulate consequences of PCSFT in terms of the 
conventional wave function. In general, deviation from Born's rule is rather small. We found an experiment  amplifying
this deviation. We start with a toy example in section 2. Then we present a more realistic example based on Gaussian 
states with very small dispersion, see section 3.}

\section{Introduction}

Recently \cite{KH0}, \cite{KH1}, a new model with hidden variables of the wave type was elaborated, so called 
{\it prequantum classical statistical field theory} (PCSFT). Roughly speaking PCSFT is a classical signal theory 
applied to a special class of signals -- ``quantum systems''. PCSFT reproduces successfully all 
probabilistic predictions of QM, including correlations for entangled systems. Moreover, PCSFT describes 
``prequantum world'' and deviations of the quantum model from  prequantum reality.
We do not want to go into details, see \cite{KH1}. We are lucky that final answers given by PCSFT in many important cases
can be reformulated by using the symbols of the conventional mathematical formalism of quantum mechanics. Thus
experimenters working in quantum foundations can proceed without studying even basic notions of PCSFT.

By PCSFT Born's rule is violated due to the presence of nonquadratic  nonlinearities in the 
process of detection (QM describes only quadratic terms). The simplest nonquadratic nonlinearity
which is taken into account by PCSFT is of the fourth order (third order nonlinearities do not produce
any statistical effect, if the prequantum random field is of the Gaussian type -- and we proceed with Gaussian fields).

Take a quantum wave function (for one dimensional system) $\Psi(x).$ Then by Born's rule the probability to find 
a system in an interval $I$ of the real line is given by
\begin{equation}
\label{BR}
{\bf p}(x \in I)= \int_I \vert \Psi(x)\Vert^2 dx.
\end{equation}

PCSFT predicts appearance of an additional term which contains the contribution of the type
\begin{equation}
\label{BR4}
\int_I \vert \Psi(x)\vert^4 dx.
\end{equation}
In fact, the situation is a little bit more complicated and the precise form of the deviation from
Born's rule will be presented later. Now we would like to discuss another important issue of the model.
The fourth order term (\ref{BR4})   contributes to the deviation with some coefficient $\alpha> 0,$ 
the dispersion of prequantum fluctuations. PCSFT does not provide a numerical value of this parameter 
of the model nor its magnitude.  Therefore the experiment should be performed for such 
quantum states, wave functions, that the contribution of the term (\ref{BR4}) will be large enough.

Of course, if G. `t Hooft \cite{H1}--\cite{H3} was right and prequantum model works only on the Planck scale, then 
the scale $\alpha$ of fluctuations of the prequantum field would be very small (of the magnitude of the 
Planck time). However, I am not as pessimistic as he and I hope that the scale of prequantum fluctuations
is not so fine. Therefore the contribution of fourth order nonlinearities may be strong enough to compensate 
smallness.

\section{Deviation from Born's rule for fourth order nonlinearities in detection}

By PCSFT, QM describes the contribution of quadratic nonlinearities in the process of detection. This 
``quadratic contribution'' is the main term in detection probabilities. By PCSFT detectors can also 
take into account nonlinearities of higher orders, the simplest one is of the fourth order. The later contribute
with a small parameter $\alpha;$ therefore their contribution is not visible in modern 
experiments which are not clean enough. That is why the quantum formalism matches so well with experimental statistical data.

We now present the results of calculation in the PCSFT framework on the  effect of the fourth order nonlinearity 
in detection. The calculations, see \cite{KH1}, are not so tricky, but they are based on integration over the 
$L_2$-space of classical prequantum fields. 

For a pure state given by the wave function $\Psi(x),$ the deviation from the basic quantum probabilistic law,
Born's rule, is approximately given by 
\begin{equation}
\label{GI} \Delta (I, \Psi, \alpha)=\alpha \Big[\int_I|\Psi(x)|^4 dx - \int_I|\Psi(x)|^2
dx \int_{{\bf R}^3}|\Psi(x)|^4 dx\Big].
\end{equation}
Thus ``generalized Born's rule'' which takes into account nonlinear fourth order effects in detection can be written as
\begin{equation}
\label{GIg} {\bf p}(x \in I) \approx  \int_I \vert \Psi(x)\Vert^2 dx + \Delta (I, \Psi, \alpha).
\end{equation}
The main difficulty is the presence of the small parameter $\alpha,$ the dispersion of fluctuations
of prequantum random fields. It is clear that it is quite small, otherwise Born's rule would be violated
long ago.

Suppose\footnote{To be mathematically rigorous, we consider $\Psi \in L_{2, 4}({\bf R}^3):$ both integrals
$\int |\phi(x)|^2 dx$ and $\int |\phi(x)|^4 dx$ are finite.} that $\rm{supp}\; \Psi \subset I,$ so the wave function is zero outside
the set $I.$ Then $\Delta \equiv 0.$ In particular,
\begin{equation}
\label{GIg1} {\bf p}(x \in {\bf R}) =  \int_{{\bf R}} \vert \Psi(x)\Vert^2 dx= 1.
\end{equation}

Let now $\Psi(x)=H, L/2 \leq x \leq L/2.$ Thus $ H^2 L=1,$ so $L=1/H^2.$ We choose $I=[0,
L/2]:$
$$
\int_I|\Psi(x)|^2 dx =1/2, \int_{{\bf R}^3}|\Psi(x)|^4 dx= H^4
L=H^2,
$$
and
$$
\int_I|\Psi(x)|^4 dx=\frac{H^4 L}{2}=\frac{H^2}{2}, \;
\Delta=\alpha (\frac{H^2}{2} - \frac{H^2}{2})=0.
$$
This calculation gave a hint that an {\it asymmetric probability
distribution} may induce  nontrivial $\Delta.$ 

We choose
\[\Psi(x)= \left\{ \begin{array}{ll}
H, \; -L/2 \leq x \leq 0\\
kH, \; 0 < x \leq L/2
 \end{array}
 \right .
\]
Hence, $1=||\Psi||^2= L H^2 (k^2 + 1)/2,$  so $L= 2/(H^2 (k^2 +
1)).$ Here $I=[L/2, 0], \int_I|\Psi (x)|^2 dx=H^2 L/2= 1/(k^2 +
1);$
$$
\int_{{\bf R}^3}|\Psi(x)|^4 dx=
%\frac{H^4 L}{2} + \frac{k^4 H^4 L}{2}=
\Big (\frac{1 + k^4}{1 + k^2} \Big ) H^2, \; \int_I|\Psi(x)|^4
dx=\frac{H^2}{k^2 + 1}.
$$
$$
\Delta=
%\alpha H^2 \Big[\frac{1}{k^2 + 1} - \frac{(1 + k^4)}{(k^2 + 1)^2}\Big]
%= \frac{\alpha H^2}{k^2 + 1} [1-\frac{1 + k^4}{1 +
%k^2}]= \frac{\alpha H^2}{1 + k^2} \big (\frac{k^2 - k^4}{1 + k^2}\big )=
\frac{\alpha H^2 k^2 (1-k^2)}{(1 + k^2)^2}.
$$
If $k > 1,$ then $\Delta(I, \Psi, \alpha) < 0.$ Suppose that $H$
increases (and $k$ is fixed) then the deviation from Born's rule
will be always negative and this deviation will be increasing. So,
for large $H$, the probability to find a system in $I$ will be
essentially less than predicted by QM. For example, choose $k=2,$
then
$$
\Delta=
%\frac{\alpha H^2 4 (-3)}{25}=
 0,48\alpha H^2.
$$
On the other hand, by choosing $k < 1,$ we shall get the positive
deviation. For $k=1,$ we have  $\Delta=0$ and there will be no
deviation from Born's rule.

The concrete form of the wave function inducing nontrivial violation of Born's rule is not important.
There are many other possibilities to make $\Delta$ large enough by taking into account 
behavior of $\vert \Psi(x)\vert^4$ on the segment $I.$ 

\section{Violation of Born's rule for Gaussian states}

Consider a Gaussian state
\begin{equation}
\label{GS}
\Psi(x)= \frac{1} {(2\pi b)^{\frac{1}{4}}} e^{- \frac{x^2}{4b} + ikx}.
\end{equation}
We select the interval 
$I=[-L/2, l/2]$ for some $L>0$ and consider the following integrals:
\begin{equation}
\label{GS1}
c_1= \int_{-L/2}^{L/2} \vert \Psi(x)\vert^2 d x 
= \frac{1}{\sqrt{2\pi b}} \int_{-L/2}^{L/2}  e^{- \frac{x^2}{2 b}} dx =
\frac{1}{\sqrt{\pi b}} \int_{-L/2\sqrt{2}}^{L/2\sqrt{2}} e^{- \frac{x^2}{b}} dx;
\end{equation}
\begin{equation}
\label{GS2}
c_2= \int_{-L/2}^{L/2} \vert \Psi(x) \vert^4 d x 
= \frac{1}{2\pi b}  \int_{-L/2}^{L/2} e^{- \frac{x^2}{b}}d x;
\end{equation}
\begin{equation}
\label{GS3}
c_3= \int_{-\infty}^{+\infty} \vert \Psi(x)\vert^4 d x 
=  \frac{1}{2\pi b}  \int_{-\infty}^{+\infty}e^{- \frac{x^2}{b}} dx=\frac{1}{2\sqrt{\pi b}}.
\end{equation}

Born's rule gives the probability to find a particle in the interval $I:$
\begin{equation}
\label{GS0}
{\bf p} (x \in I)= c_1.
\end{equation}
Our prequantum model predicts deviation from this probability; this deviation 
is approximately equal to
$$
\Delta(I, \Psi, \alpha) \equiv \Delta(L, b, \alpha) =
\alpha[c_2 - c_1 c_3]=\frac{\alpha}{2 \pi b}[   \int_{-L/2}^{L/2} e^{- \frac{x^2}{b}}d x - 
 \int_{-L/2\sqrt{2}}^{L/2\sqrt{2}} e^{- \frac{x^2}{b}} dx].
 $$
 Thus 
\begin{equation}
\label{GSQ}
\Delta(L, b, \alpha)= \frac{\alpha}{\pi b} \int_{-L/2\sqrt{2}}^{L/2} e^{- \frac{x^2}{b}} dx.
\end{equation}
For a fixed state $\Psi,$ we are interested in approaching the maximal deviation from 
Born's rule. We shall see that deviation is maximal for some special $L$ depending on the dispersion
of the Gaussian state. 

We have
$$\frac{\partial \Delta(L, b, \alpha)}{\partial L}= \frac{\alpha}{2\sqrt{2}\pi b} \left[ \sqrt{2}
e^{- \frac{L^2}{4b}} - e^{- \frac{L^2}{8b}}\right]=0.
$$
Then 
\begin{equation}
\label{GSQ1}
L_{\rm{max}}= 2 \sqrt{b \ln 2}.
\end{equation}
We can easiy check that this is the point of maximum and that 
\begin{equation}
\label{GSQ2}
\Delta(L_{\rm{max}}, b, \alpha)= \frac{\alpha}{\pi \sqrt{b}} 
\int_{\sqrt{\frac{\ln 2}{2}}}^{\sqrt{\ln 2}} e^{- x^2} dx.
\end{equation}
Set 
$$
\gamma = \frac{1}{\pi} \int_{\sqrt{\frac{\ln 2}{2}}}^{\sqrt{\ln 2}} e^{- x^2} dx.
$$
Then
\begin{equation}
\label{GSQ3}
\Delta_{\rm{max}}=  \gamma \frac{\alpha}{\sqrt{b}}.
\end{equation}
Let 
$$
\alpha \sim 10^{-m},
$$
where $m$ is sufficiently large. Then, to get deviation of the magnitude $\sim 10^{-s},$ we should be able
to prepare a Gaussian state with the dispersion 
$$
b \sim  10^{-2m + 2s}
$$

\medskip

{\bf Conclusion.} {\it We presented the experimental design which might induce violation of Born's rule
due to nonlinear (fourth order) effects in detection. To perform experiments of this kind, one should be 
able to play with preparation of pure states (for a single particle). One possibility is to prepare 
Gaussian states with very small dispersion. Successful realization of this experiment
will be definitely a great new step in creation of a proper description of microworld.}

\section{A hint from PCSFT}

By PCSFT ``quantum particles'' are symbols used to denote classical random signals -- classical fields (electromagnetic for photon,
``electronic'' for electron) fluctuating on a fine (prequantum) time scale. A position detector performs 
spatial integration of such a prequantum signal $x\to \phi(x).$ The main contribution is given by the quadratic term;
this is Born's rule. However, a detector integrates not only quadratic nonlinearity, but even nonlinearities 
of higher orders. The simplest one integrates the following functional  of the prequantum field:
\begin{equation}
\label{PVG}
\pi_{2,4}(\phi)= \vert \phi(x)\vert^2 +  \alpha \vert \phi(x)\vert^4,
\end{equation}
where $\alpha> 0$ is a small parameter.

\medskip

 I would like to thank H. Rauch, G.`t Hooft, S. Stenholm,  A. Zeilinger, G. Weihs, F. De Martini, M. Genovese, M. D´Ariano,
V. Manko and O. Manko for stimulating conversations on (im)possibility
to go beyond QM.

\end{document}